\documentclass[aps,preprint,amsmath,amssymb]{revtex4}
\usepackage{graphicx}
\begin{document}

\begin{center}
{\large\bf Iron pnictide superconductors: Electrons on the verge}
\\[0.5cm]

Qimiao Si
\\

{\em Department of Physics and Astronomy, Rice University,
Houston, TX 77005, USA}\\

\end{center}

\vspace{0.5cm}
{\bf
An optical analysis reveals that the electronic correlations in the
`parent' compounds of the iron pnictide superconductors are sufficiently
strong to significantly impede 
the mobility of the electrons.
}

%\newpage 
\vspace{0.5cm}

Superconductivity occurs in diverse materials. Until 1986, the highest
observed superconducting transition temperature ($T_c$) was 23 K, 
prompting some researchers to suggest that the conventional mechanism
based on electron-phonon coupling is incapable of yielding transition
temperatures much higher. Then came the discovery of the copper oxide
superconductors, which, after two decades, still hold the record 
$T_c$ of over 130 K at ambient pressure. The recent 
discovery \cite{Kamihara:JACS08} of superconductivity at 26 K 
in LaFeAsO$_{\rm 1-x}$F$_{\rm x}$ has generated
much activity, and $T_c$ has since been pushed to 
over 55 K in related FeAs-based materials \cite{Zhao_Sm1111}. 
Like the copper oxides, the iron pnictides have well-defined parent
systems, such as LaFeAsO and BaFe$_{\rm 2}$As$_{\rm 2}$.
Typically, superconductivity arises when charge carriers (either electrons
or holes) are introduced into the parent compound, for example 
by substituting fluorine for oxygen in LaFeAsO. Fluorine donates its extra
electron to the FeAs layer, 
thereby changing the layer's carrier concentration. 

Conventional superconductors are made from simple metals. Their electrons 
are weakly correlated, with mutual interactions that are small in comparison
with their kinetic energy; it suffices to consider such essentially free
electrons coupled to
% vibrating ions, or 
phonons. At the opposite limit,
copper oxide superconductors involve strongly correlated electrons. 
The repulsive interaction among the electrons is so strong that the 
parent compounds are `Mott insulators'; energetically speaking,
it pays for the electrons 
to stay localized. The question of how strong the electron correlations
are in the iron pnictide superconductors is as important as it is 
delicate. On page 647 of this issue \cite{Qazilbash08}, 
Mumtaz Qazilbash et al. 
determine the strength of these correlations using optical measurements
of the kinetic energy of the electrons.

The correlation question is not merely academic, as different answers 
lead to different points of departure for the study of the low-energy 
physics of magnetism and superconductivity. Early clues were ambiguous.
The parent iron arsenides are antiferromagnetic, which suggests that
some amount of electronic correlation must be present. However, the parents
are always metallic, indicating that the correlations fall short 
of the Mott limit, 
in contrast to the copper oxides.
%which would result in an antiferromagnetic insulator
%(as for the cuprates).

One perspective is that the parent iron arsenides are weakly correlated, 
and that the antiferromagnetism arises in the way it does in the standard 
itinerant antiferromagnet, chromium. This picture invokes a strong
`nesting' of the Fermi surfaces, which provides an enhanced phase
space for exchange interactions among the well-defined electronic
states near the Fermi surfaces, and leads to an antiferromagnetic
ground state even though $U/t$,
the ratio of the characteristic Coulomb repulsion, $U$, and the kinetic
energy, or bandwidth, $t$,
might be relatively small. 

%%%% Figure 1%%%%
\begin{figure}[htbp]
   \centering
   \includegraphics[width=7in]{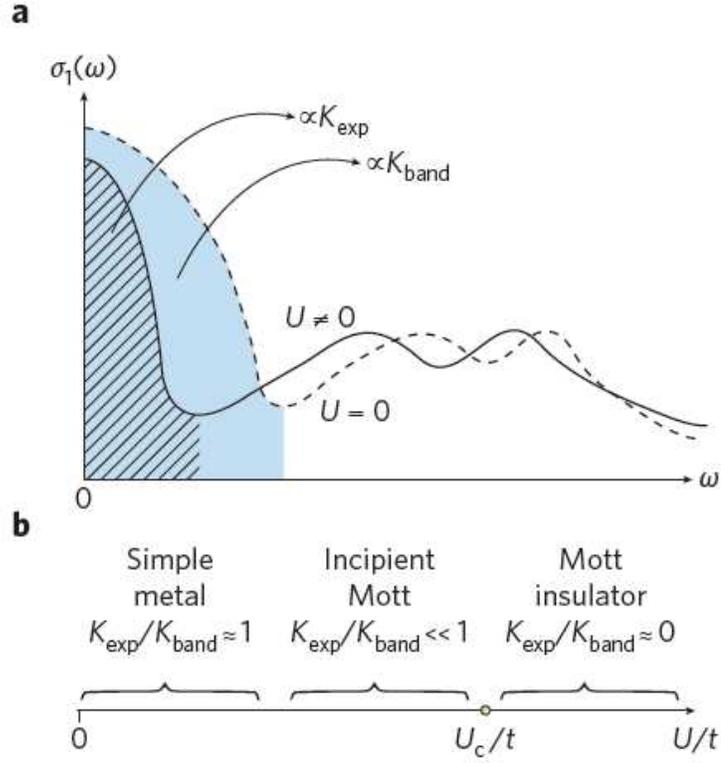}
   \caption[]
  {Electronic correlations in the undoped iron pnictides. 
{\bf a.} Optical conductivity as a function of frequency. 
The area under the $\omega=0$ Drude peak is 
(for certain simplified energy dispersions) proportional to the kinetic 
energy of the coherent electrons near the Fermi energy, yielding 
$K_{\rm band}$ and $K_{\rm exp}$ for the non-interacting and interacting cases,
respectively. The degree to which $K_{\rm exp}/K_{\rm band}$ is smaller than 
one is a measure of the interaction-induced reduction of the electrons' 
mobility and, by extension, the normalized electron-electron 
repulsion, $U/t$. The interaction transfers spectral weight from 
low energies to high energies, up to order $U$. {\bf b.} Different 
regimes of correlation. $U_c$ is the threshold interaction for a Mott 
localization transition in a system with an integer number of electrons
partially occupying some electronic bands. That $K_{\rm exp}/K_{\rm band}$
is substantially smaller than one means that $U/t$ is 
smaller than but close to $U_c/t$, where the electrons 
are on the verge of losing their mobility.
}
   \label{fig:drude_incipient_mott}
\end{figure}
%%%%%%%%%%%%%

An alternative view is that the electronic correlations in the parent
iron arsenides, although somewhat weaker than those in the copper oxides,
are still sufficiently strong to place the system close to the boundary
between itinerancy and interaction-induced electronic localization.
In other words, they are `bad metal' states in proximity to 
a Mott transition.

The work of Qazilbash and co-workers \cite{Qazilbash08}
provides insight into this issue. The authors measure the optical
conductivities of LaFePO, the phosphorus counterpart of the arsenic-based
LaFeAsO, and ${\rm BaFe_2As_2}$. They compare their room-temperature
results with other  electronic systems of varying degrees of correlation: 
parent and doped copper oxides; transition-metal compounds, such as 
${\rm V_2O_3}$, which are known to be bad metals close to Mott transition;
and relatively simple metals such as copper, silver, chromium and 
${\rm MgB_2}$. They analyse the quantity
\begin{eqnarray}
K_{\rm exp} = \frac{2\hbar^2 c}{\pi e^2} \int d \omega 
\sigma_{\rm Drude} (\omega) , 
\nonumber
\label{kinetic-energy}
\end{eqnarray}
where $\sigma_{\rm Drude}$ is the measured Drude part of the optical 
conductivity, $\omega$ denotes frequency, and $c$ is the inter-FeAs-layer
distance.
This quantity has the physical meaning of the kinetic
energy associated with the coherent part of the single-electron 
excitations in an interacting metallic system. It can be compared
to $K_{\rm band}$, which is the kinetic energy of the underlying 
non-interacting system, and is in practice extracted from 
{\it ab initio} band-structure calculations. Coulomb repulsion
impedes itinerancy and, hence, renders $K_{\rm exp}$ smaller than 
$K_{\rm band}$. The ratio $K_{\rm exp} / K_{\rm band}$ therefore provides 
a measure of the degree of correlation, as shown in Figure 1.
It should take the value one in a non-interacting electron system; 
for a Mott insulator, it should be zero at zero temperature
and remain a small number, much less than one, at room temperature.

Qazilbash et al. show that the value of $K_{\rm exp} / K_{\rm band}$ at
room temperature is indeed close to one for simple metals and 
is considerably smaller than one for a Mott insulator. The value
for the parent iron arsenide, ${\rm BaFe_2As_2}$,
is about $0.3$ -- the same order of magnitude 
as the optimally doped copper oxides or the incipient Mott insulator 
${\rm V_2O_3}$. For the parent iron phosphide, LaFePO, the value 
is about $0.45$, which is somewhat larger than 
%the iron arsenide 
for ${\rm BaFe_2As_2}$
but still not close to one.

One natural interpretation of these results is that the parent iron 
arsenides are indeed located near the boundary between itinerancy 
and localization. Further support for this interpretation can 
be inferred from the optical conductivity measurements of several
other groups \cite{Hu08,Yang08,Boris09}, which have shown that
changes in temperature can induce transfer of optical spectral 
weight between the low-energy part of the spectrum and the part 
at high energies, 
%over 
greater than $1$ eV. Such a spectral-weight transfer 
is a hallmark of bad metals close to Mott localization, which 
feature incoherent electronic excitations in the form 
of `precursor Hubbard bands' (Figure 1) at some distance 
from the Fermi energy.

To fully establish that the incipient Mott localization operates in the 
iron pnictides, the precursor Hubbard bands need to be probed. One difficulty
lies in there being several $d$ bands, for the precursor Hubbard bands can 
be embedded in other high-energy features associated with interband 
transitions. (This makes the temperature-induced spectral-weight 
transfer mentioned above especially illuminating.) In addition,
it remains to be seen whether the system can be tuned across the Mott
transition and into a Mott insulator state. The observations 
of Qazilbash et al. in fact provide a clue to this. The phosphides are 
shown to be less correlated than the arsenides, presumably because they
have a larger chemical pressure (as the phosphorous ion is appreciably
smaller). Will the replacement of arsenic by antimony or bismuth give rise 
to Mott insulators?

Despite these cautionary considerations, it is tempting to draw implications
of the strong-coupling picture for low-energy physics. The existence 
of substantial incoherent
weight in the excitation spectrum makes it meaningful to consider nearly
localized magnetic moments, coupled by superexchange interactions, 
which have been shown to lead to the observed magnetic structure. 
This strong-coupling approach seems to provide a good basis 
for understanding the magnetic dynamics as well. Inelastic neutron 
scattering experiments \cite{zhao,Diallo}
have recently identified spin-wave-like excitations all the way 
to the antiferromagnetic zone boundary. The spin waves have 
a fairly high energy ($\sim 200$ meV) that, incidentally, 
bodes well for a magnetic mechanism for superconductivity.
They also have a very large spectral weight.

Superconductivity comes further down the energy hierarchy. 
The observation of sizeable electronic correlations supports
strong-coupling approaches to superconducting pairings. 
More generally, it supports the widely held belief that 
the mechanism for superconductivity in the iron pnictides 
lies in electron-electron interactions, and not in the 
standard electron-phonon coupling.

\end{document}